\documentclass[aps,prb,twocolumn,superscriptaddress,floatfix]{revtex4}
\usepackage{amsmath}
\usepackage{amssymb}
\usepackage{graphicx,graphics}
\usepackage{latexsym,verbatim}
\usepackage{dcolumn} 
\usepackage{color}
\usepackage{cancel}
\usepackage{ulem}

\begin{document}

\title{Impact of electrostatic fields in layered crystalline BCS superconductors}

\author{Luca Chirolli}
\affiliation{Department of Physics, University of California, Berkeley, CA-94720}
\affiliation{Instituto Nanoscienze-CNR, I-56127 Pisa, Italy}

\author{Tommaso Cea}
\affiliation{Instituto de Ciencia de Materiales de Madrid, CSIC, ES-28049 Cantoblanco Madrid, Spain}
\affiliation{Imdea Nanoscience, ES-28049 Cantoblanco Madrid, Spain}

\author{Francesco Giazotto}
\affiliation{Instituto Nanoscienze-CNR, I-56127 Pisa, Italy}

\begin{abstract}
Motivated by recent experiments reporting the suppression of the critical current in superconducting Dayem bridges by the application of strong electrostatic fields, in this work we study the impact on the superconducting gap of charge redistribution in response to an applied electric field in thin crystalline metals. By numerically solving the BCS gap equation and the Poisson equation in a fully self-consistent way, we find that by reducing the pairing strength we observe an increased sensitivity of the gap on the applied field, showing sudden rises and falls that are compatible with surface modifications of the local density of states. The effect is washed out by increasing the pairing strength towards the weak-to-moderate coupling limit or by  introduction  of a weak smearing in the density of states, showing the evolution from a clean crystal to a weakly disordered metal.
\end{abstract}

\maketitle

\section{Introduction} 
\label{Sec:Intro}

A thorough understanding of the impact of electrostatic fields on superconductivity is of great relevance from the fundamental point of view, as it may represent an easy-access active knob to control fundamental quantum states of matter and open the way to a number of technological applications. Electrostatic fields have been successfully employed in systems where the carrier density is low, such as thin crystalline films \cite{shi2015superconductivity,leng2012indications,shiogai2016electric,hendrickx2018gate-controlled}, band insulators  \cite{ueno2008electric,ueno2011discovery,ye2012superconducting}, interfaces \cite{caviglia2008electric}, semiconducting and low-density two-dimensional materials \cite{saito2015metallic,li2016controlling,sajadi2018gate-induced,chen2019signatures}, or to tune the proximity effect \cite{morpurgo1999gate-controlled,jarillo-herrero2006quantum,tsuneta2007gate-controlled}. In most of the cases the electrostatic field controls the carrier density by shifting the active bands of the material. A low electronic density yields  poor screening and a sufficiently thin structure ensure full penetration of the electric field.  In contrast, metals characterized by a high carrier density screen very well the electrostatic field within few Angstroms from the surface \cite{PinesNozieres,AshcroftMermin,Mahan,GiulianiVignale} and the residual skin contribution is typically negligible at the level of carrier density and density of states.  Superconductors realized in diffusive metals are typically hardly affected by electrostatic fields. 

Recently, a series of experiments conducted on metallic superconducting Dayem nanobridges have shown that a strong electric field, generated by a high voltage applied on a side gate, is able to switch the critical current $I_c$ of the Dayem bridge off in a reversible ambipolar way \cite{desimoni2018metallic,paolucci2018ultra,paolucci2019magnetotransport,paolucci2019field,desimoni2020niobium}. Although a certain material dependence is observed, with a pronounced effect in Nb, Va, Ti, the effect seams to be relatively general, as it occurs in  Al-Cu-Al proximity Josephson junctions \cite{desimoni2019josephson} and in  Al Dayem bridges \cite{bours2020unveiling}. Possible leakage currents and related overheating effects  \cite{Golokolenov2020ontheorigin,alegria2020high-energy,ritter2020superconducting} have been minimized by constructing suspended nanobridges \cite{rocci2020suspended}. Switching current distribution measurements show the presence of very strong gate-induced phase fluctuations \cite{puglia2020phase-slip}. Theoretical attempts to explain the origin of the observed field effect suggested a surface orbital polarization \cite{mercaldo2020electrically,bours2020unveiling,fukaya2020orbital}  and a possible Schwinger effect \cite{solinas2020schwinger}.  These findings are yet to be understood in the usual framework of the BCS theory and represent a challenge from the fundamental point of view, whose solution may reveal of  great technological interest. Although the observed phenomenology involves diffusive polycrystalline metals, it offer the opportunity to study the impact of an electric field in systems at the boundary between low-density crystalline materials and diffusive metals. 

In this work we address the problem of gate-controlling superconductivity in thin metallic clean systems, consisting in a crystal composed by $N$ layers and characterized by a large carrier density, large DoS at the Fermi level, but with still a well defined notion of discreteness. Rather then studying the effect of variation of the carrier density, we focus on the impact of a redistribution of charge in response to an applied electric field by choosing an anti-symmetric profile of the potential in a capacitor-like configuration, that guarantees only the bipolar part of the effect is described, leaving aside global carrier density modifications. We numerically solve the fully self-consistent gap equation and Poisson electrostatic equation describing simultaneous condensation of a superconducting gap and screening of the applied field within the BCS theory. The screening length is only slightly increased from its metallic counterpart, in agreement with random-phase approximation results first described by Anderson \cite{anderson1958rpa} and Thouless \cite{thouless1960perturbation}, that predict a correction of order $(\Delta/E_F)^2$  [\onlinecite{Mahan}],  with $\Delta$ the superconducting gap and $E_F$ the Fermi energy \cite{virtanen2019superconducting}. 

In order to enhance the responsiveness of the system, we choose as the basic system a tight-binding model in the cubic lattice, but results are qualitatively confirmed with an in-plane triangular lattice. We find that for a small relative dielectric constant, relatively large system size, and strong superconducting pairing, the gap is essentially insensitive to the applied field. In turn, by reducing the pairing strength we observe an increased sensitivity to the applied field, with the gap showing sudden rises and falls as the applied voltage is increased. This behavior originates from the density of states  modification induced by the screened potential. The latter adds to the confining potential  and plays the role of a layer chemical potential. Although the exponentially decaying profile significantly modifies only the outermost few layers, our results show that it can result in sizable bulk effects. For a perfectly clean crystalline structure the entire density of states spanning the whole bandwidth is necessary to account for the observed behavior. In turn, the introduction of a weak energy smearing in the DoS, that emulates the effect of weak disorder, washes out the effect and the gap follows mainly the DoS at the Fermi level. 

Our results apply to weak coupling limit and show how an evolution from a clean to a weakly disordered system takes place. These results are expected to be significant for layered materials and  thin cristalline metals and predicts a certain degree of control of superconductivity an applied electrostatic field.

\section{Mean-field BCS with screening}
\label{Sec:1}
 
\begin{figure}[t]
\includegraphics[width=0.45\textwidth]{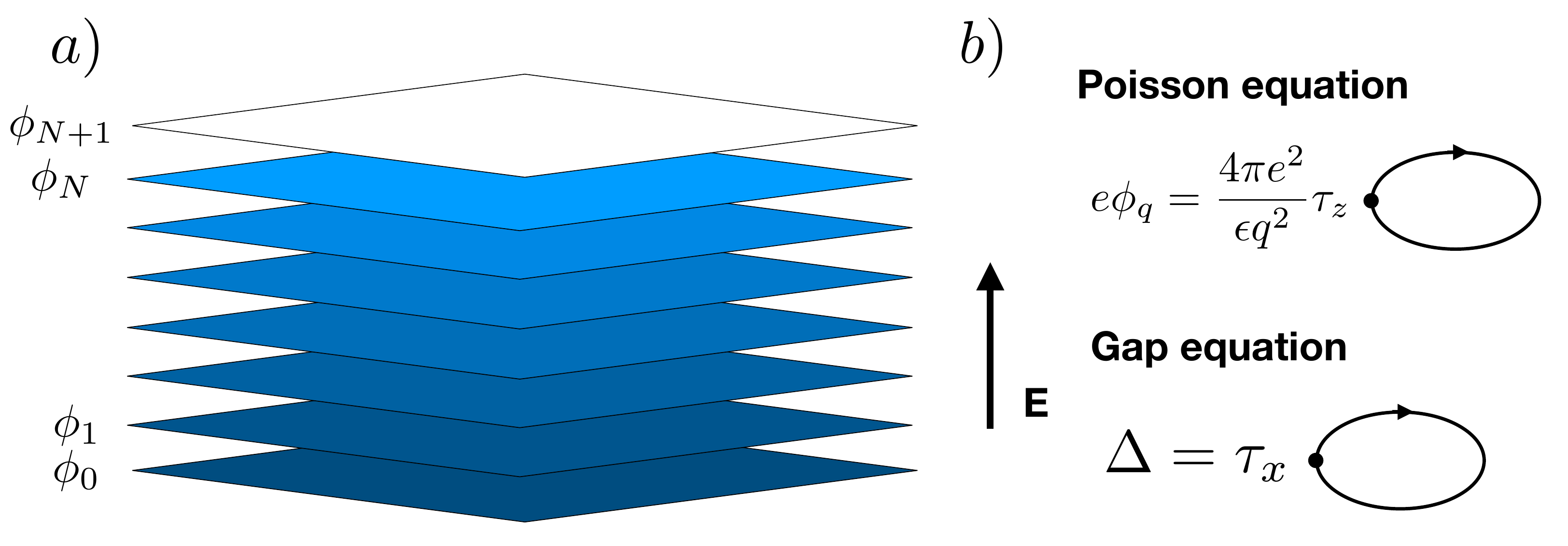}
\caption{a) Schematics of the system composed by $N$ layers. The electric field is applied along the out-of-plane direction and it is modeled by fixing the electrostatic potential at the $i=0$ and $i=N+1$ layers. b) Diagrammatic form of the Poisson equation and the gap equation. 
\label{Fig:1}}
\end{figure}

We consider a system composed by $N$ layers, as depicted in Fig.~\ref{Fig:1}, described each by a spin-degenerate microscopic tight-binding model. For simplicity we assume a single orbital per unit cell, with nearest neighbor hopping $t$, that results in a dispersion $\epsilon_{\bf k}$, with ${\bf k}$ in-plane momentum. Interlayer nearest neighbor hopping is described by the same hopping $t$. Electrons interact via a purely local two-body attraction described by a Hubbard term with strength $U$, that acts as a pairing interaction. In addition, we apply an external electric field ${\bf E}_g$ along the out-of-plane $z$ direction via a side gate. The latter is screened by the electron gas and the full electric field $E^z_i$ can be introduced, that is described via an electrostatic energy potential $\phi_i$, such that $E^z_{i+1}=-(\phi_{i+1}-\phi_i)/ae$, with $a$ the lattice constant and $e$ the electric charge. The full Hamiltonian then reads
\begin{eqnarray}\label{FullH}
H&=&\sum_{{\bf k},i,s}(\epsilon_{\bf k}-\phi_i)c^\dag_{{\bf k},i,s}c_{{\bf k},i,s}-t\sum_{i,s}c^\dag_{{\bf k},i+1,s}c_{{\bf k},i,s}+{\rm H.c.}\nonumber\\
&-&U\sum_{{\bf k},{\bf k}'}c^\dag_{{\bf k},\uparrow}c^\dag_{-{\bf k},\downarrow}c_{-{\bf k}',\downarrow}c_{{\bf k}',\uparrow}+V_C,
\end{eqnarray}
where $V_C$ is the Coulomb interaction. The latter is crucial to correctly describe screening of the applied field. Its continuum form in  Fourier transform is given by $V_C({\bf q},q_z)=4\pi e^2/(\epsilon(q^2+q_z^2))$, with $\epsilon$ the dielectric coupling constant. The momentum transfer $({\bf q},q_z)$ is restricted to non zero value to account for the background positive charge. We separate the in-plane and out-of-plane Coulomb interaction by singling out the ${\bf q}=0$, $q_z\neq0$ term, that reads
\begin{equation}\label{Eq:VCeff}
V_C=\frac{1}{2}\int dzdz'n(z)V_{\rm eff}(z-z')n(z'),
\end{equation}
with $V_{\rm eff}(z-z')=\frac{1}{L}\sum_{q_z\neq 0}e^{iq_z(z-z')}4\pi e^2/(\epsilon q_z^2)$,  $n(z_i)=\sum_{{\bf k},s}c^\dag_{{\bf k},i,s}c_{{\bf k},i,s}$, and neglect the residual interaction. Standard decoupling of Eq.~(\ref{Eq:VCeff}) gives rise to the Poisson equation. By noticing that $q_z^2$ is the eigenvalue of the Laplacian in 1D, we directly write a discrete version of the Poisson equation in 1D as
\begin{equation}\label{Eq:Poisson}
-\phi_{i+1}-\phi_{i-1}+2\phi_i=\frac{4\pi e^2}{\epsilon a}\left[n_i-n_0\right],
\end{equation}
that ensures charge conservation locally on each layer. 

The rest of the Hamiltonian is decoupled in the Cooper channel at mean-field level by defining a local layer dependent gap, $\Delta_i=\langle c_{-{\bf k},i,\downarrow}c_{{\bf k},i,\uparrow}\rangle$.  The Bogoliubov deGennes tight-binding Hamiltonian is written as
\begin{equation}\label{Eq:BdG}
H=\sum_{{\bf k},ij}h_{ij}({\bf k})c^\dag_{{\bf k},i,s}c_{{\bf k},j,s}+\Delta_ic^\dag_{{\bf k},i,\uparrow}c^\dag_{-{\bf k},i,\downarrow}+{\rm H.c.}
\end{equation}
with $h_{ij}({\bf k})=[\epsilon_{\bf k}-\phi_i]\delta_{ij}-t(\delta_{i+1,j}+\delta_{i,j+1})$ and the chemical potential has been absorbed in the dispersion. The gap and the electron density are then written as
\begin{equation}\label{Eq:GapCharge}
n_i=\frac{2}{N_k}\sum_{\bf k}v^2_i({\bf k}),\qquad 
\Delta_i=\frac{U}{N_k}\sum_{\bf k}u_i({\bf k})v_i({\bf k}).
\end{equation}
where the $u_i({\bf k})$ and $v_i({\bf k})$ are the particle and hole part of the eigenvectors at layer $i$ of the BdG Hamiltonian Eq.~(\ref{Eq:BdG}) and $N_k$ is the number of ${\bf k}$-points in the BZ. 

The equation for the gap and the density (\ref{Eq:GapCharge}) are solved iteratively together with the Poisson equation (\ref{Eq:Poisson}). For simplicity we assume to insert the layered system in a capacitor like structure, that fixes the value of the gate voltage to be opposite on the two sides of the system. A uniform gap $\Delta^{(0)}$ and a linear potential $\phi^{(0)}_i=(2i/N-1) \phi_g$ are assigned  at the first step and the charge density and the gap are recursively updated. The potential $\phi_i$ is obtained via inversion of the Poisson equation (\ref{Eq:Poisson}) and by imposing a fixed boundary condition at the two most external layers $\phi_0=-\phi_{N+1}=\phi_g$. The mean density $n_0$ is kept fixed at half-filling and the chemical potential is updated to keep the half-filling condition at every step. Convergence is achieved within a threshold error smaller than $10^{-5}$ of a chi-squared error function for the three quantities $n_i,\Delta_i,\phi_i$. By increasing the gate in a discretized way, convergence speed highly increases using as guesses for the  gap, density, and potential those self-consistently obtained at a smaller value of the gate potential.

\section{Density of States and Screening}
\label{Sec:2}

\begin{figure}[t]
\includegraphics[width=0.45\textwidth]{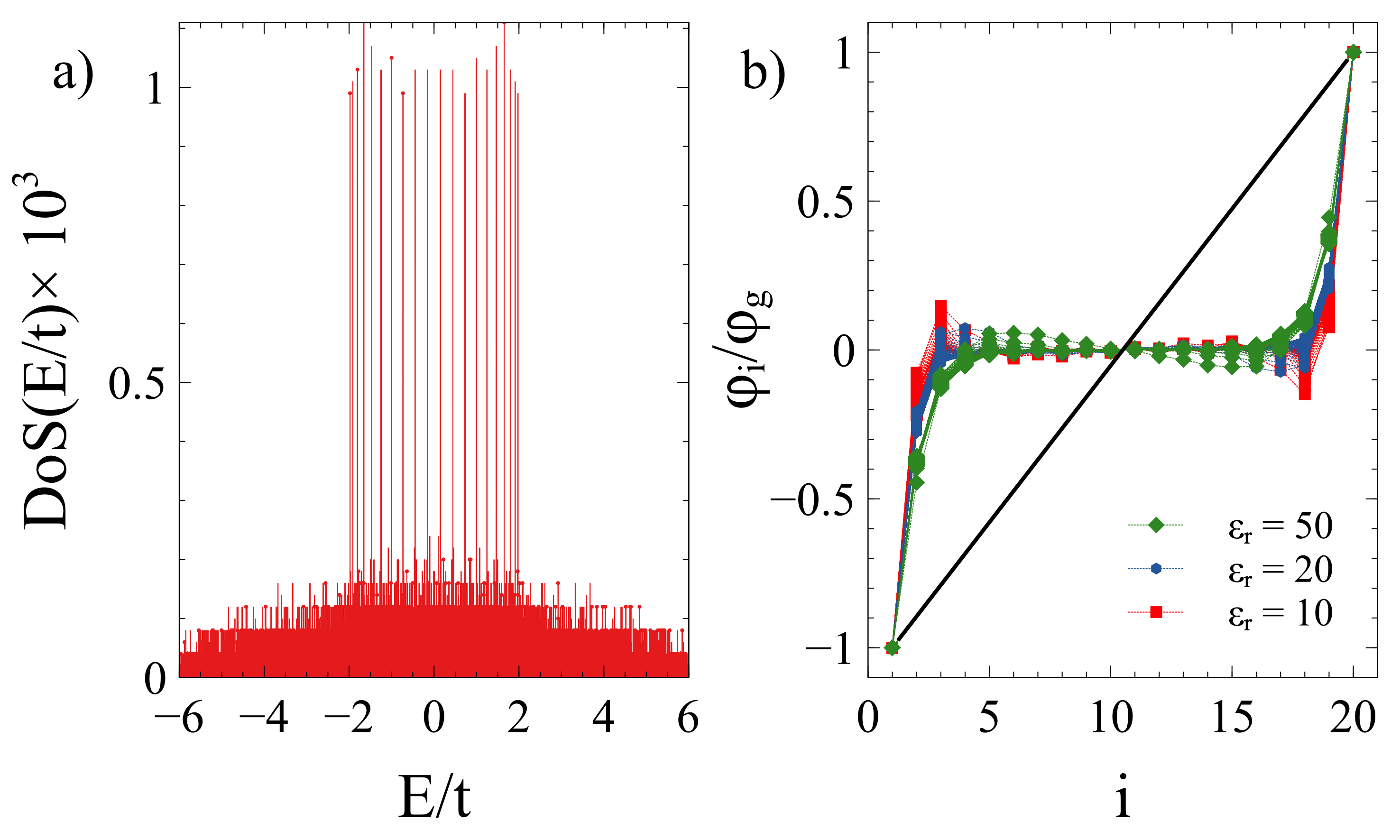}
\caption{a) Histogram of the eigenvalues multiplicity (total density of states) in absence of applied gate. b) Profile of the self-consistent potential for different values of the relative dielectric constant $\epsilon_r=10, 20, 50$.  
\label{Fig:2}}
\end{figure}

We assume the in-plane system to be described by a square lattice, so that the dispersion reads 
\begin{equation}
\epsilon_{\bf k}=-2t(\cos(k_x)+\cos(k_y)). 
\end{equation}
At half filling the chemical potential is $\mu=0$. For a structure composed by a few layers some concerns may arise from the half-filling van Hove singularity that characterizes the one-layer density of states. Strictly speaking, if $N$ is odd a smeared van Hove singularity still appears at half filling, whereas for $N$ even it is shifted at positive and negative energies by inter-layer tunneling.  In order to avoid peaks in the DoS at the Fermi level we always choose an even number of layers. 

We measure the strength of the Hubbard attraction $U$ in units of the hopping energy $t$ and group the lattice spacing $a$ and the dielectric constant $\epsilon$ in the energy scale $e^2/(a\epsilon)$. Setting $a=1$~\AA, we define our Rydberg as $Ry=e^2/(\epsilon_0a)=18.05~{\rm eV}\equiv 18.05~ t$. This way, the only free parameters in the system are the number of layers $N$, the attraction strength $U$, the number $N_k=N_xN_y$ of ${\bf k}$-points in the BZ and the relative dielectric constant $\epsilon_r$. The smaller is the dielectric constant $\epsilon_r$ the strongest is the screening.

In Fig.~\ref{Fig:2}a) we show a histogram of the multiplicity of the eigenvalues of a $N_x=N_y=100$ grid in momentum space, for a system constituted by $N=20$ layers. The histogram tends to the typical DoS of a 3D cubic lattice once a smearing in energy is assumed. Deviations due to finite size effects both in-plane and out-of-plane are evident and $N=20$ peaks reminiscent of the original van Hove singularities are still present. 

\begin{figure}[t]
\includegraphics[width=0.45\textwidth]{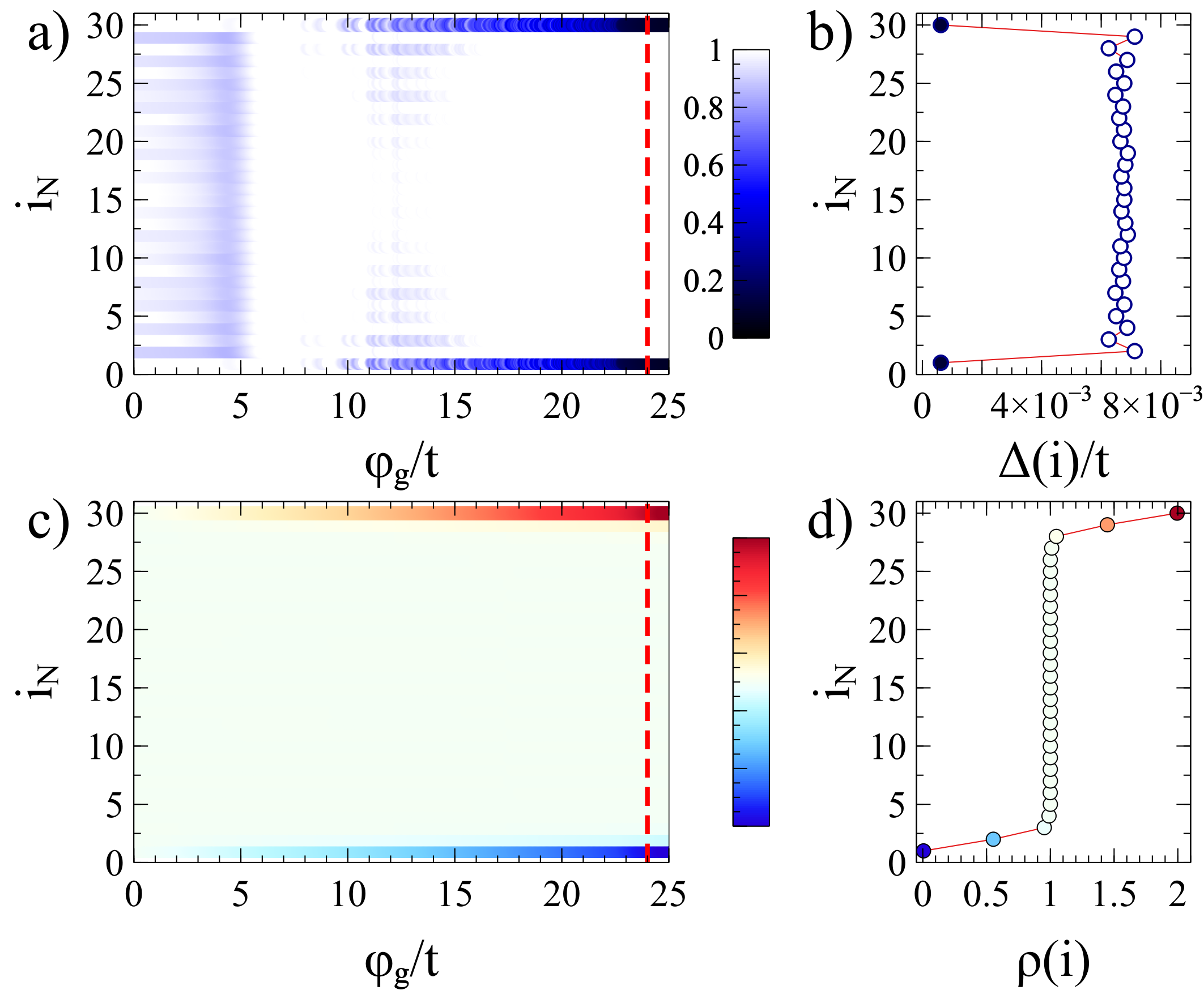}
\caption{Results of the simulations for a strong pairing $U=1~t$ for a thick slab with $N=30$. a) Color plot of the local gap versus layer $i_N$ and applied gate voltage $\phi_g$.  b) Section of a) at $\phi_g=24~t$ showing gap suppression in the outer most layers. c) Value of the self-consistent charge density versus the layer $i_N$ and the applied gate, showing charge depletion in the outer most layers. d) Self-consistent charge density at $\phi_g=24~t$ showing full depletion at the outermost layer.
\label{Fig:S1}}
\end{figure}

In Fig.~\ref{Fig:2}b) we show the self consistent potential normalized to its maximum strength for values of the applied gate $0<\phi_g<5t$, for three different values of the relative dielectric constant $\epsilon_r=10, 20, 50$. The field is screened very well by the metal and an overall exponential decay is recognized. Besides, oscillations of the charge on the scale of the screening length are present,  indicating a local increase/decrease of the layer chemical potential $\phi_i$ in the bulk of the slab. The charge distribution screens the external field via charge accumulation at the outermost layers. The latter overshoots the one necessary to screen the field and a series of alternating electric dipoles on the scale of the screening lengths are generated to compensate local overshooting. 

\section{Simulations}

We now present the results of full numerical calculations. As pointed out in Sec.~\ref{Sec:Intro}, we expect no effect of the applied gate for a large system, characterized by strong screening and a relatively strong pairing, within the weak coupling limit. We confirm this expectation for a system composed by $N=30$ layers, pairing strength $U=1~t$, relative dielectric constant $\epsilon_r=10$ and  $N_x=N_y=100$ points in momentum space. The results of the simulations are shown in Fig.~\ref{Fig:S1}. The self-consistent gap and potential are shown in Fig.~\ref{Fig:S1}a) and c), respectively. The gap is mostly uniform through the slab for all values of the applied electric field. Small oscillations on the scale of the lattice constant appear on top of the average value. For strong field the gap in the outermost layer approaches zero at $\phi_g=24~t$. This arise because of complete charge depletion/saturation in the outermost layer (shown in Fig.~\ref{Fig:S1}d) and the absence of available particle (hole) states locally suppresses the gap.

\begin{figure}[t]
\includegraphics[width=0.45\textwidth]{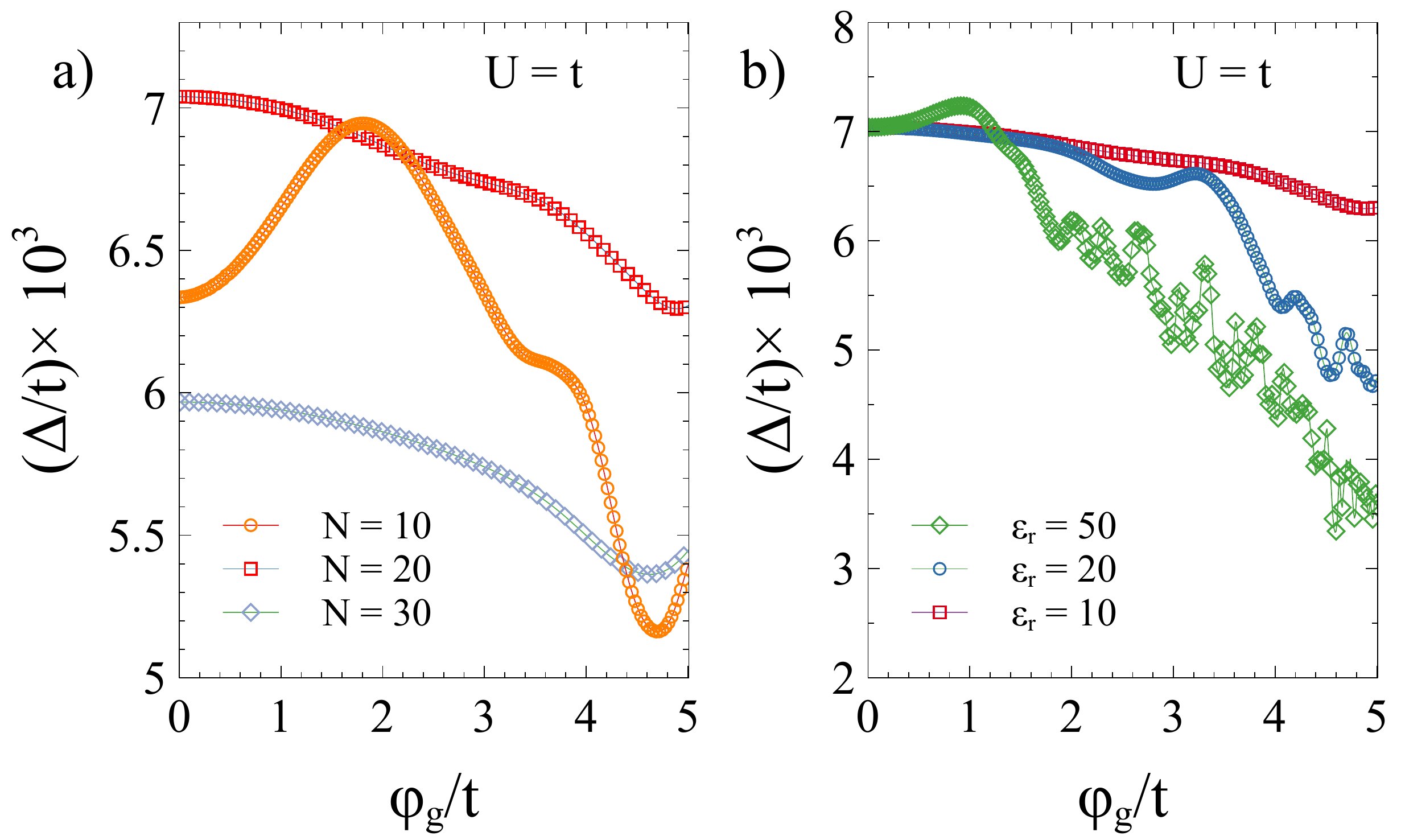}
\caption{Average gap $\Delta$ versus applied gate $\phi_g$ for $U=t$  varying a) the number of layers $N=10,20,30$ for $\epsilon_{r}=10$, and b) the relative dielectric constant $\epsilon_r=10, 20, 50$ for $N=20$. In-plane grid with $N_x=N_y=100$. 
\label{Fig:4}}
\end{figure}

{\bf Average gap varying $N$ and $\epsilon_{r}$.---}
We now study the impact of the applied field on the gap for samples with different thickness $N=30, 20, 10$,  for  $U=t$. Results are shown in Fig.~\ref{Fig:4}a). The gap at zero applied gate voltage is not a monotonic function of the sample thickness and shows well know quantum oscillations. The dependence on the applied gate is very smooth for $N=20, 30$ and the gap shows robustness to the applied electric field. For $N=10$ a sizable and smooth modulation of the gap is observed.  We then fix the slab thickness to $N=20$ and study the dependence of the layer-averaged gap $\Delta=\sum_i\Delta_i/N$ on the applied field for three values of the dielectric constant $\epsilon_{r} = 10, 20 , 50$. Results are shown in Fig.~\ref{Fig:4}b). By increasing the value of $\epsilon_r$ the gap shows an average smooth linear decrease with the applied gate, that seams to saturate for higher values of $\epsilon_r$, accompanied by fast irregular fluctuations on top of the linear decrease. The analysis is repeated for $U=0.8~t$ (results not shown). The average gap becomes more sensitive to the applied gate and strongly non-monotonic rises and falls appear in correspondence of the smooth variations observed for $U=t$.

\begin{figure}[t]
\includegraphics[width=0.45\textwidth]{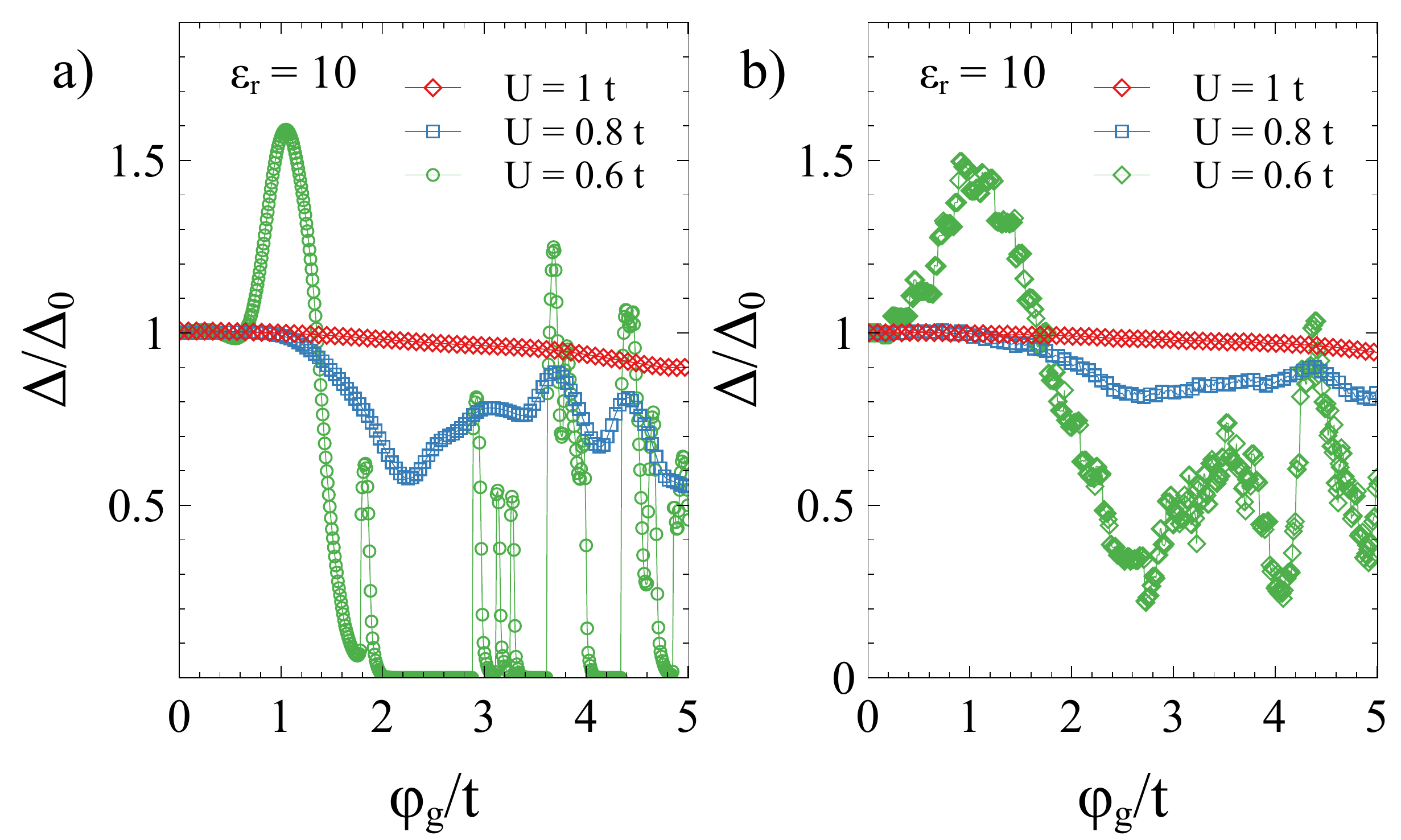}
\caption{a) Normalized average gap versus applied gate voltage $\phi_g$ for different values of the pairing strength $U/t = 1, 0.8, 0.6$ for  $\epsilon_r=10$. b)  Gap calculated through Eq.~(\ref{Eq:GapEqBCS}) with the density of states $\nu(E)$ resulting from the self-consistent potential. Fixed parameters are $N=20$ and $N_x=N_y=100$. 
\label{Fig:5}}
\end{figure}

{\bf Average gap varying $U$.---}
We now study the gap versus gate voltage curve varying $U$. The zero voltage gap clearly diminishes by decreasing $U$, so that we normalize the curves with their zero voltage value $\Delta_0=\Delta(\phi_g=0)$. In Fig.~\ref{Fig:5}a) we show the results for  $\epsilon_r=10$. We see that by decreasing $U$ the effect of the gate becomes more and more pronounced. For $U=0.6$ at $\phi_g\simeq 1.5~t$ the gap drops to zero after a 50\% increase and it stays zero a part from very sharp and sudden revivals. For the chosen value $\epsilon_r=10$ the field penetrates only for few layers. The weak modulation of the gap observed for large $U$ is strongly amplified for smaller $U$. The analysis is repeated for $\epsilon_{r}=50$ (results not shown) and the dependence on the gate voltage becomes more and more frustrated. We observe again that the small fluctuations appearing in the $U=t$ curves are strongly amplified for smaller $U$, suggesting a common density of states origin.

\section{Gap from DoS}

The analysis so far presented shows that a strong sensitivity to the gate appears when reducing the pairing strength $U$. We point out that in all simulations a smooth convergence of the self-consistent calculation is observed at every steps, ruling out numerical instability. The BCS theory predicts that all properties of the superconducting gap are determined by the DoS of the system. The latter is typically assumed to be uniform over a large range of energies where the pairing is active and approximated with its value at the Fermi level. Clearly, this approximation fails if the DoS suffers strong modifications close to the Fermi level due to confinement, as shown in Fig.~\ref{Fig:2}a), where the DoS at the Fermi level is ill-defined.

We then calculate the gap that results assuming the effect of the electric field comes solely by the screened potential $\phi_i$. This is done by taking the self-consistently screened potential for two given $\epsilon_r=10, 50$ for every value of the applied gate voltage, calculating the DoS $\nu(E)$ in the normal phase, and self-consistently solving for the gap via the BCS gap equation
\begin{equation}\label{Eq:GapEqBCS}
1=U\int dE \frac{\nu(E)}{2\sqrt{E^2+\Delta^2}},
\end{equation}
by varying only $U$. The result is shown in Fig.~\ref{Fig:5}b): there is a striking similarity between the curves shown in Fig.~\ref{Fig:5}a), showing how the unexpected behavior of the gap is totally understood in terms of the full DoS and its readjustment with the screened potential. Although the dependence on the field is smoother in the curve in Fig.~\ref{Fig:5}b), the rises and falls appear in the same ranges of gate voltage.

\begin{figure}[t]
\includegraphics[width=0.45\textwidth]{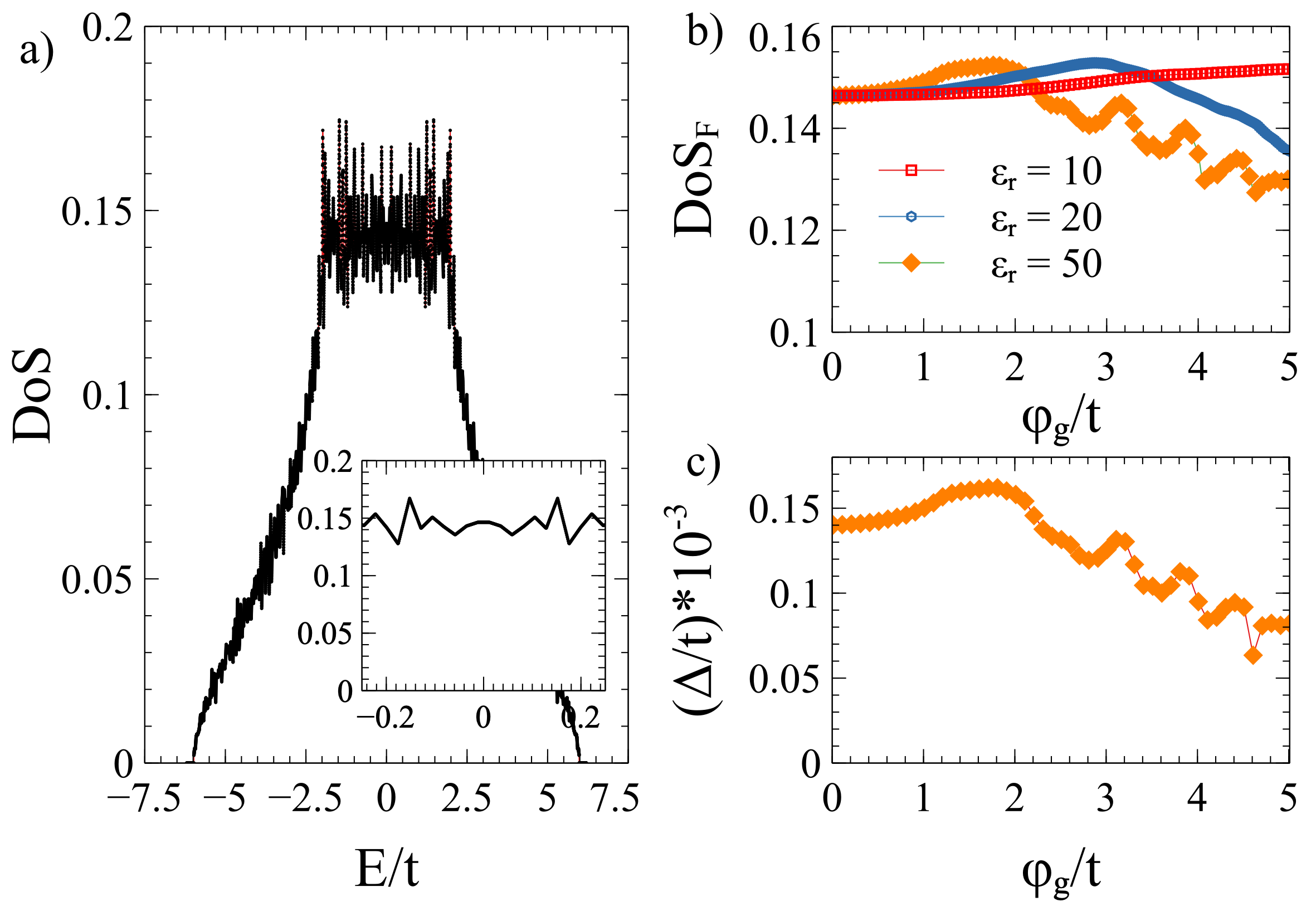}
\caption{a) Density of states versus energy after a smearing with broadening $10^{-5}t$ for a system with $N=20$ layers and $\epsilon_r=50$ at zero applied field. Inset: zoom on a small energy window around the Fermi level at $E=0$. b) DoS at the Fermi level as a function of the applied gate voltage for $\epsilon_r=10, 20, 50$. c) Gap resulting from solution of Eq.~(\ref{Eq:GapEqBCS}) for $\epsilon_r=50$ and $U=0.6 t$. 
\label{Fig:6}}
\end{figure}

\section{Extrapolations}

It is clear that finite size effects, in particular the finite grid in momentum space, are at the origin of the observed sensitivity to the applied gate voltage for weaker value of $U$. We can mimic a denser grid in momentum space and a small amount of local disorder by smearing the DoS on a small energy window. We notice that it is sufficient to introduce a very small broadening on order $10^{-5}t$ to smear the DoS without completely erasing its discrete origin, as shown in Fig.~\ref{Fig:6}a). The DoS at the Fermi level is now a meaningful quantity and it is shown in Fig.~\ref{Fig:6}b) for three values of $\epsilon_r=10, 20, 50$. We can then calculate the self-consistent gap via numerically integrating Eq.~(\ref{Eq:GapEqBCS}) for $U=0.6 t$ and $\epsilon_r=50$. We see that the gap closely follows the behavior of the DoS at the Fermi level. This result reliably predicts a modulation of the gap in a weakly disordered metal via an external electric field that penetrates sufficiently in the system. It also shows how the sensitivity to the applied gate evolves from clean thin crystal to a weakly disordered metal. Furthermore, comparison of Fig.~\ref{Fig:6}c) with Fig.~\ref{Fig:4}b) for $\epsilon_{r}=50$ shows how the gap versus  gate voltage for small $U$ and weak energy smearing is compatible with the gap dependence at large $U$ without energy smearing. This way, the results for weakly disordered weak-coupling limit are analogous to those of moderate-strong pairing in a clean system.

\section{Small grains and dot}

\begin{figure}[t]
\includegraphics[width=0.45\textwidth]{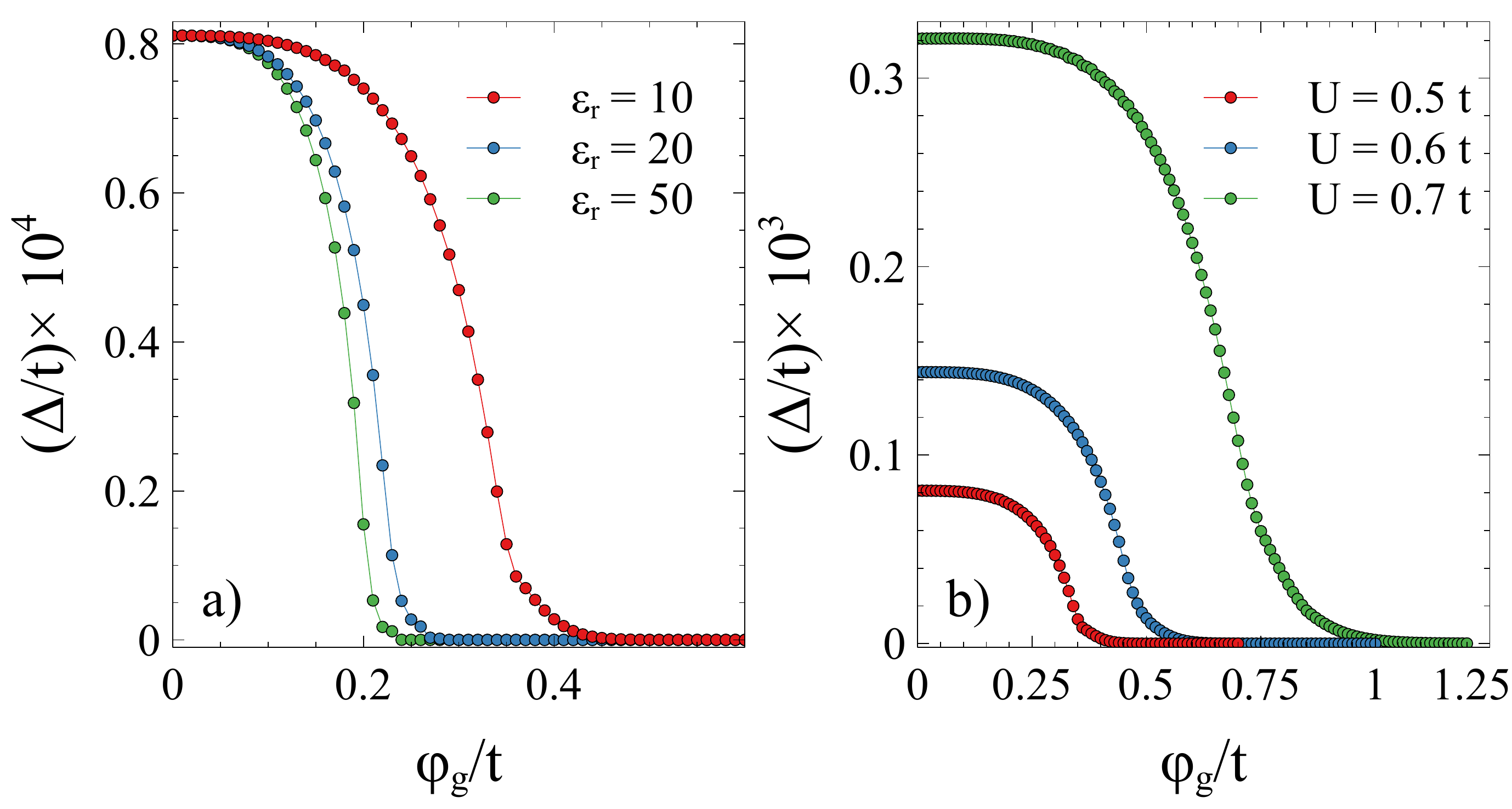}
\caption{Gap as a function of the applied gate voltage for a system characterized by $N=20$ and  $N_x=N_y=80$, with in-plane periodic boundary conditions. a) $U=0.5~ t$ varying $\epsilon_r$.  b) $\epsilon_r=10$ varying $U$. 
\label{Fig:7}}
\end{figure}

Finally, we study a smaller system that rather describes a clean dot. This is done by keeping the number of layers to $N=20$ and reducing the size of the in-plane lattice by setting $N_x=N_y=80$. This results in a reduced DoS, with consequent reduction of the gap size at zero gate voltage with respect to the cases analyzed in the previous sections. The average gap $\Delta$ versus applied gate is shown in Fig.~\ref{Fig:7} and smoothly decreases to zero, in a fashion similar to the BCS temperature dependence.  Fig.~\ref{Fig:7}a) we fix $U=0.5~t$ and vary the relative dielectric constant $\epsilon_r$.  The curves fall all on top of each other upon proper field rescaling (not shown). In Fig.~\ref{Fig:7}b) we increase the pairing strength and confirm that for a larger gap the result remains valid.

\section{Discussion}
\label{Sec:3}

\begin{figure}[t]
\includegraphics[width=0.45\textwidth]{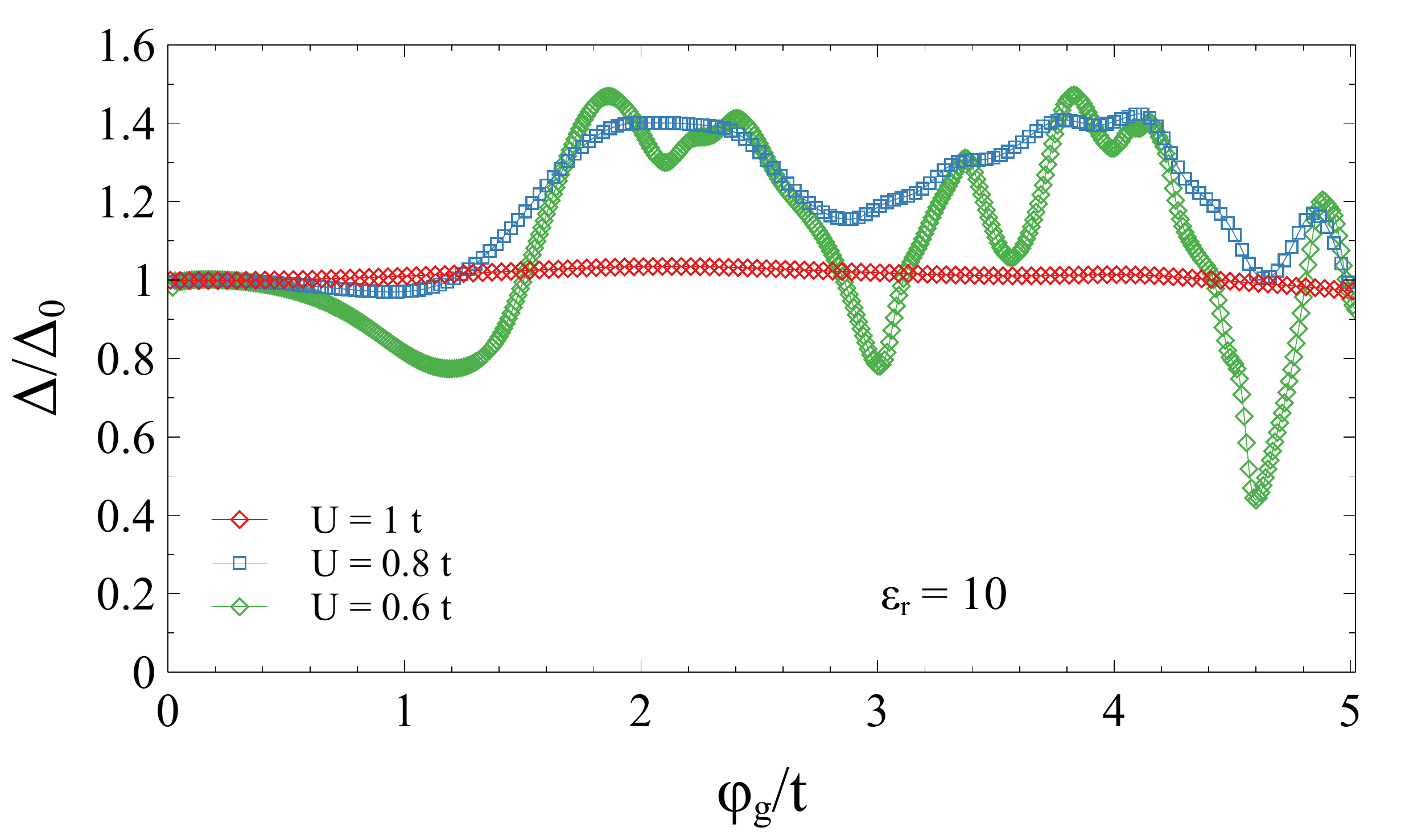}
\caption{Gap as a function of the applied gate potential for a in-plane triangular tight-binding model characterized by $N=20$,  $\epsilon_r=10$ and $N_x=N_y=100$ for different values of the pairing $U/t=0.8, 0.6, 0.5$. 
\label{Fig:8}}
\end{figure}

The analysis presented tackles the problem of the simultaneous condensation of a superconducting gap and screening of an externally applied field in a fully self-consistent way. The Poisson equation (\ref{Eq:Poisson}) includes the part of the Coulomb interaction that describes repulsion of the average charge of each layer and results from minimization of the total energy. It does not include a finite in-plane momentum transfer. The problem is solved exactly at the mean-field level and does not account for phase fluctuations. 

The results presented show that the behavior of the system is totally due to modification of the density of state induced by the gate voltage, that acts as a confining potential. The observations are qualitatively confirmed by changing the in-plane lattice model. It is well known that the square lattice at half filling represents somewhat a peculiar case, with van Hove singularities close to the Fermi level. In Fig.~\ref{Fig:8} we show the results of the simulation for an in-plane triangular lattice. The electric field is screened in few lattice constants, depending on the relative dielectric constant $\epsilon_r$. Fluctuations of the gap are seen for reduced strength of the pairing $U$, confirming the results obtained for the square lattice. The exponential dependence of the gap on the density of states enhances weak variation of the latter when reducing the pairing strength $U$. Differently from the case of larger in-plane systems, the results are not confirmed for a small system, for which the gap is never suppressed.  

The results and methodology open the way to reliable modeling of systems where the surface physics has a non-trivial content, such as the case of a Rashba spin-orbit interaction, that is controlled by an applied electric field, or a multi-orbital character of the band structure, that enables an electric-field controlled orbital-Rashba effect. Screening in absence of pairing converges very quickly, and the joint impact of electric field, sample geometry and Rashba field can show non-trivial results on the gap.   

In summary, we find that in cristalline thin metals a strong sensitivity to an applied electric field appears in the weak coupling limit, with the gap showing sudden rises and falls as the applied voltage is increased. This behavior reflects the density of states modification induced by the screened potential acting mostly on the outermost few layers. For a perfectly clean crystalline structure the observed behavior can be understood only in terms of the entire density of states spanning the whole bandwidth. The introduction of a weak energy smearing in the DoS emulates the effect of weak disorder and washes out the effect, showing a gap that follows mainly the DoS at the Fermi level. Our results are expected to be significant for layered materials and thin cristalline metals allowing control of superconductivity by an externally applied electrostatic field.

\section{Acknowledgments}

L.C. acknowledges the European Commission for funding through the MSCA Global Fellowship grant TOPOCIRCUS-841894. T.C. acknowledges fundings from the European Commission, under the Graphene Flagship, Core 3, grant number 881603. F.G.  acknowledges the EU’s Horizon 2020 research and innovation program under grant agreement No. 800923 (SUPERTED), and  the European Research Council under the EU’s Horizon 2020 Grant Agreement No. 899315-TERASEC for partial financial support.

\bibliography{Bibfile}{}

\appendix

\end{document}